\documentclass[10pt,a4paper,twoside]{article}
\usepackage{epsfig}
\usepackage{baltlat6}
\usepackage{array}
\usepackage{here}

\begin{document}
\ \
\vspace{0.5mm}
\setcounter{page}{277}
\vspace{8mm}

\titlehead{Baltic Astronomy, vol.\,xx, xx--xx, 2012}

\titleb{SPECTRAL MONITORING OF AGNs:\\
 PRELIMINARY RESULTS FOR Ark 564 AND Arp 102B}

\begin{authorl}
\authorb{A. I. Shapovalova,}{1}
\authorb{L. \v {C.} Popovi\' {c},}{2}
\authorb{D. Ili\' {c},}{3}
\authorb{A. Kova\v {c}evi\'{ c},}{3}
\authorb{J. Kova\v {c}evi\'{ c},}{2}
\authorb{A. N. Burenkov,}{1}
\authorb{V. H. Chavushyan}{4}
\end{authorl}

\begin{addressl}

\addressb{1}{Special Astrophysical Observatory of the Russian AS, Nizhnij Arkhyz,
\\ Karachaevo-Cherkesia 369167, Russia; ashap@sao.ru}
\addressb{2}{Astronomical Observatory, Volgina 7,\\ 11160 Belgrade, Serbia; lpopovic@aob.bg.ac.rs}
\addressb{3}{Department of Astronomy, Faculty of Mathematics, University of Belgrade\\ Studentski trg 16, 11000 Belgrade, Serbia;}
\addressb{4}{Instituto Nacional de Astrofísica, Óptica y Electrónica, Apartado Postal 51, CP 72000, Puebla, Pue, Mexico}
\end{addressl}

\submitb{Received:   ; accepted:  }

\begin{summary}
We present preliminary results of the long term spectral monitoring
of two active galactic nuclei with different broad line shapes:  Ark
564 and Arp 102B. Ark 564 is a bright nearby narrow line Syfert 1
(NLS1) galaxy with  relatively narrow permitted optical emission
lines and a high FeII/H${\beta}$ ratio, while Arp 102B is a nearby
broad-line radio galaxy with broad double-peaked Balmer emission
lines. The spectra of Ark 564 were observed during 11-year period
(1999-2009) and the spectra of Arp 102B in the 12-year period
(1998-2009), with SAO 6-m and 1-m telescopes (Russia) and the GHAO
2.1-m telescope (Cananea, Mexico).
\end{summary}

\begin{keywords} galaxies: active / quasars: individual: Ark 564, Arp 102B / line: profiles \end{keywords}

\resthead{Spectral monitoring of AGNs:\\
 Preliminary results for ARK564 and ARP102B}
{A. I Shapovalova, L.\v C. Popovi\'c, D. Ili\'c et al.}

\sectionb{1}{INTRODUCTION}
Active galactic nuclei (AGN) often exhibit variability in the broad
emission lines. In spite of many papers devoted to the physical
properties (physics and geometry, see e.g. Sulentic et al. 2000) of
the broad line region (BLR) in AGN, the true nature of the BLR is
not completely revealed. The broad emission lines (BEL), can give us
many information about the BLR geometry and physics. Especially the
variability in the BEL profiles and intensities could be used for
investigating the BLR nature.

A long-term spectral monitoring of the nucleus of some AGN has
revealed a time lag in the response of the broad emission lines
relative to flux changes in the continuum (e.g. Wanders and Peterson
1996, Kollatschny and Dietrich 1997). This lag depends on the size,
geometry, and physical conditions of the BLR. Thus, the search for
correlations between the nuclear continuum changes and flux
variations in the broad emission lines may serve as a tool for
mapping the geometrical and dynamical structure of the BLR (see e.g.
Peterson 1993, and reference therein). During the past decade, the
study of the BLR in some objects has achieved considerable success,
mainly because of the increasing number of coordinated
multiwavelength monitoring campaigns through the international "AGN
Watch" campaign (see e.g., Peterson 1999). In several papers, we
presented results  of our long term monitoring of  NGC4151, NGC5548,
and 3C390.3 (see Table 1). However, in this paper we present the
preliminary analysis of the spectral monitoring of Ark 564 and Arp
102B during periods 1999-2009 and 1998-2009, respectively.

\begin{table}[ht!]
\begin{center}
\vbox{\footnotesize\tabcolsep=3pt
\parbox[c]{124mm}{\baselineskip=10pt
{\smallbf\ \ Table 1.}{\small\
Basic data for the selected objects: object name, monitoring period, redshift, radius of the BLR determined from the H${\beta}$ line,
spectral characteristics (DPL stands for double peaked line; CST for changing spectral type), AGN type and reference.
\lstrut}}
\resizebox{12cm}{!} {
\begin{tabular}{ccccccc}
\hline
\hline
Object& Period& z&$R_{H{\beta}}$ &Spectral&AGN type&Reference \hstrut\lstrut    \\
 name&  [years]& & [ld] & characteristics &&\hstrut\lstrut \\
\hline
NGC 4151&1996-2006&0.0033  &1-50  & CST &Sy. 1.5-1.8 & Shapovalova et al. 2008, 2010a \hstrut  \\
NGC 5548& 1996-2004& 0.0172 &6-26 &CST&Sy. 1.0-1.8 & Peterson et al. 2002, Shapovalova et al. 2004 \\
 &  &   &  & &   &Ili\'{ c} 2007, Popovi\' {c} et al. 2008\\
3C 390.3&1995-2007& 0.0561 & 35-100 &DPL&RL QSO& Shapovalova et al. 2010b, Popovi\'{ c} et al. 2011\\
      &  &   &  & &   & Jovanovi\'{ c} et al. 2010\\
Arp 102B&1998-2009& 0.0242& & DPL  &RL QSO&in prep\\
Ark 564&1999-2009& 0.0247& & strong Fe II& NLSy1 &in prep\lstrut\\
\hline
\end{tabular}
}}
\end{center}
\end{table}

Spectral monitoring  was carried out at the 6-m and 1-m telescopes
of SAO RAS and at the 2.1-m telescope of INAOE (Cananea, Mexico).
Observations were fulfilled with long-slit spectrographs equipped
with a CCD. The spectral range was 4000-8000\AA, the resolution
3-15\AA, and the S/N$>$50 in the continuum near the H$\alpha$ and
H$\beta$ lines. For details on data acquisitions, data reduction and
calibration see Shapovalova at al. (2008, 2010b).

\sectionb{2}{PRELIMINARY RESULTS FOR Ark 564 and Arp 102B}

Ark 564 was the object of one of the most intensive broad-band
reverberation mapping programs undertaken to date, which aimed to
determine the nature of the relationship between X-ray and
UV-optical continuum variations and thus obtain an estimate of the
BLR size and viral mass of the central source. Ark 564 was observed
by ASCA (Pounds et al. 2001, Edelson et al. 2002), Hubble Space
Telescope (Collier et al. 2001, Crenshaw et al. 2002) and from many
ground based observatories as part of an International AGN Watch
project (1998 Nov to 2001 Jan, Shemmer et al. 2001). Ark 564 has
shown a strong associated UV absorber (Cranshaw et al. 1999). There
are indications that it also possesses a warm X-ray absorber, as
seen by the absorption lines of OVII and OVIII detected in a Chandra
spectrum (Matsumoto et al. 2001).

An example of a Ark 564 spectrum obtained from our monitoring of this object is given in Fig. 1.
The variability of the flux is clearly seen (Fig 1, left panel).
We also fitted one spectrum of Ark 564 using Gaussian components for all
lines (see Fig 1, right panel). We fitted the region of H$\beta$ line where we have
a strong contribution of Fe II multiplet (for details on fitting procedure
see Kova\v {c}evi\'{c} et al. 2010). The best-fitting is nicely following the
observed spectrum (see the residual in Fig 1, right panel), especially the Fe II line group,
that have similar widths as the intermediate line component of the H$\beta$ line.

\begin{figure}[ht]
\centering
\includegraphics[width=6cm]{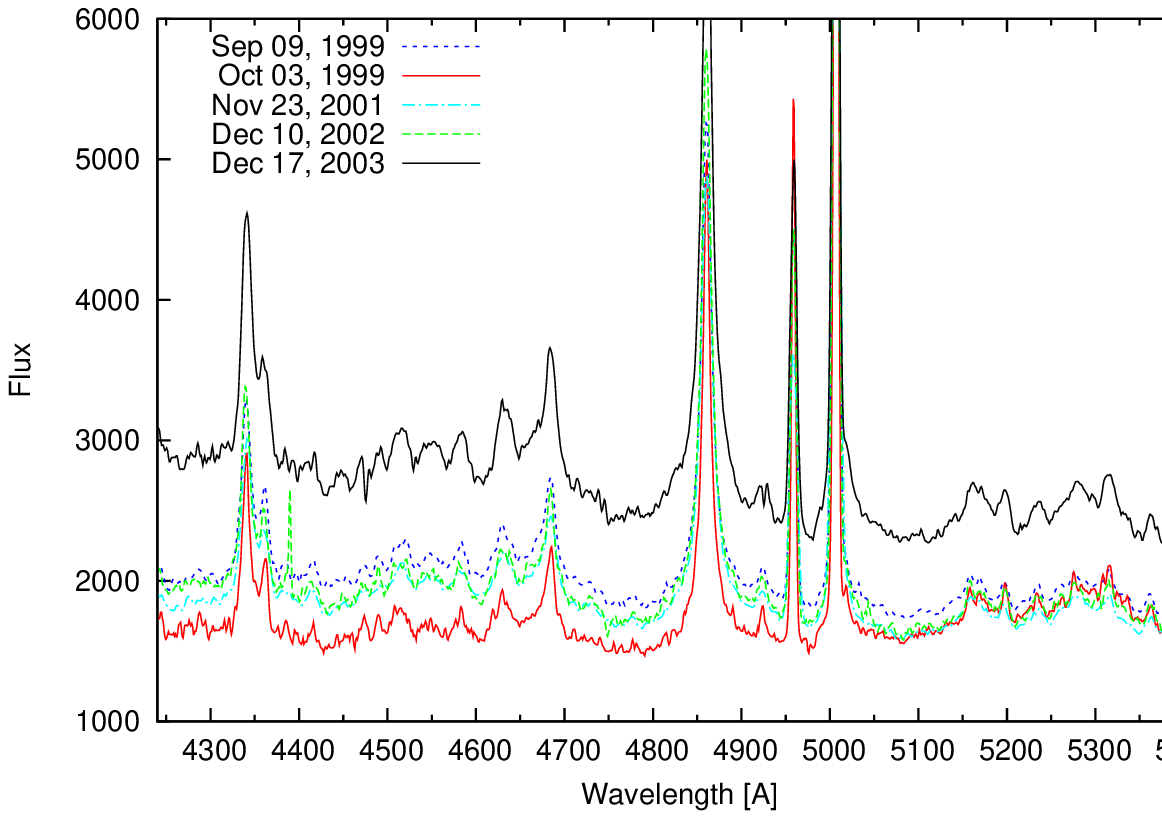}
\includegraphics[width=6cm]{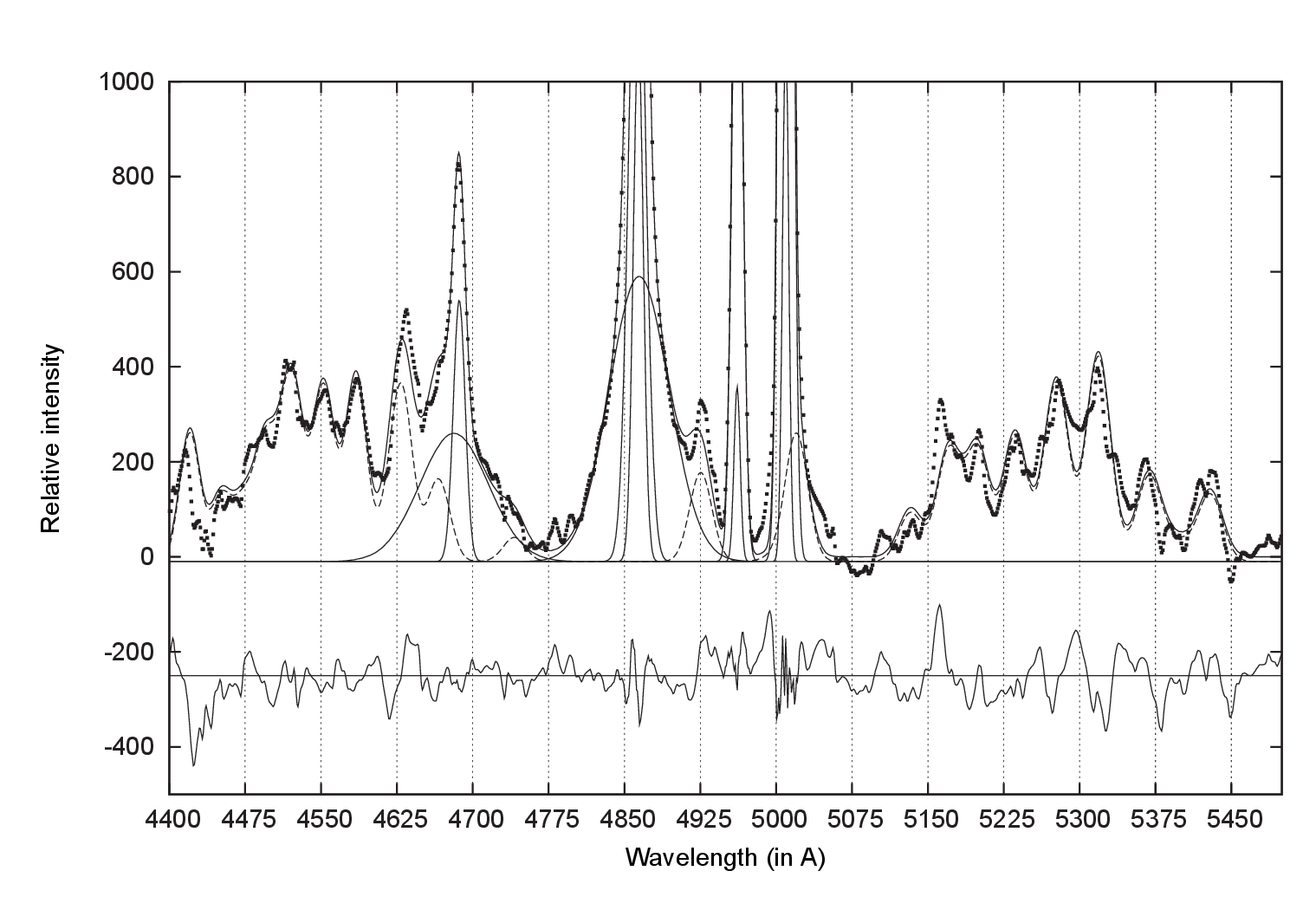}
\caption{Left panel: Flux variability of Ark 564. Dates are given at
top left corner. Right panel: An example of fitted spectrum of Ark
564 using Gaussian components for all lines.} \label{fig1}
\end{figure}

Arp 102B emits double-peaked line profiles (see Fig. 2). The first
and the simplest model of a region which can produce such line profiles 
is the model of an accretion disk or disk-like region. On the other hand,
this object has very strong low ionization lines but almost absent 
high ionization lines (Stauffer et al. 1983; Halpern et al. 1996). 
These weak high ionization lines have no double-peaked profiles,
thus can be explained with the accretion disk model only with adding more
assumptions. The observed properties of this object such as hard X-ray source,
low Eddington luminosity and no strong UV bump, indicate a model of 
an advection dominated accretion flow (Ho et al. 2000).

The shape of line profiles of the broad component of H$\alpha$ emission line
varied strongly (see Fig. 2), but in all cases there is a prominent
bump in the blue wing and less prominent one with a flat top in the
red wing of the broad H$\alpha$ component.

\begin{figure}[ht]
\centering
\includegraphics[width=8cm]{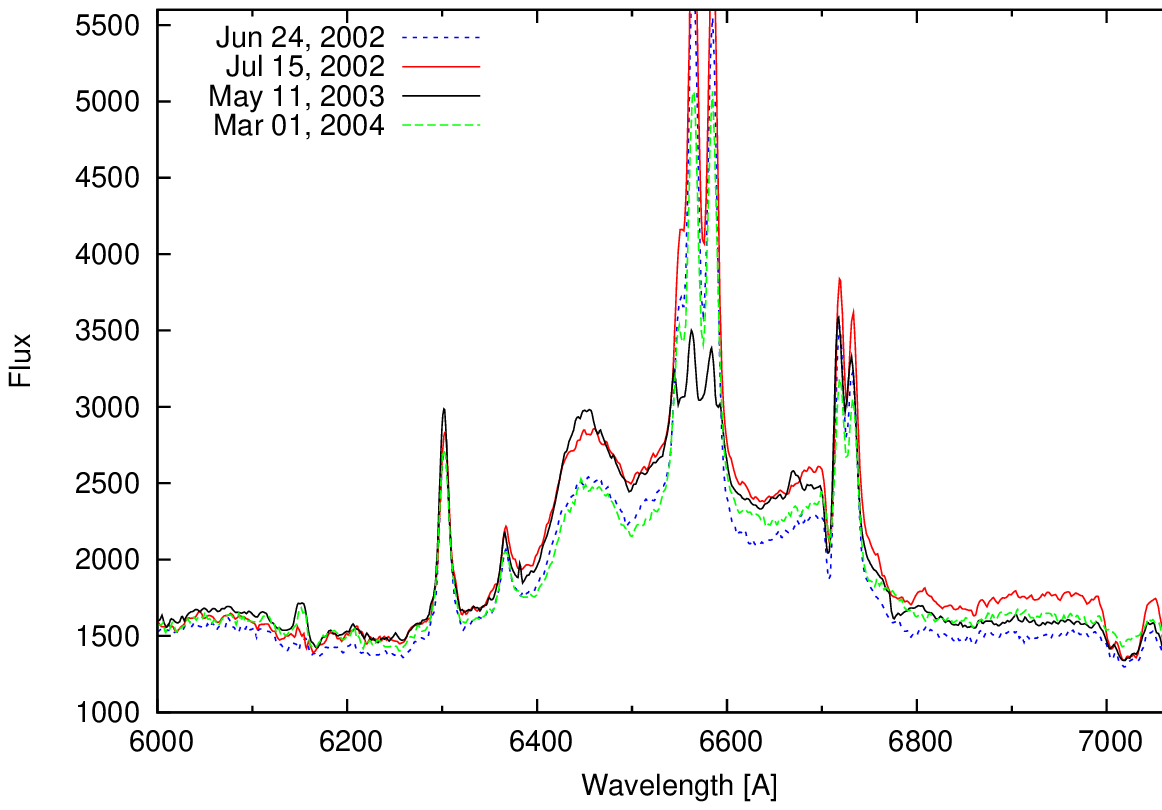}
\caption{Flux variability of Arp 102B. Dates are given at top left
corner. } \label{fig2}
\end{figure}

The broad lines of Arp 102B can be well modelled with a
relativistic disc model given by Chen and Halpern (1989). 
The best fits with this disk model give such inclination angles
that imply an intermediate viewing angle, a characteristic for a number AGNs 
with double-peaked emission line profiles (Eracleous and Halpern 1994). 
However AGNs with double-peaked broad line profiles like Arp102B, are apparently 
very rare (around 10\% of AGNs with broad lines), and they are mainly radio-loud 
objects. This should be further investigated. 

More detailed analysis and discussion about monitoring
of these two AGNs will be given elsewhere.

\thanks{RFBR (grants N97-02-17625 N00-02-16272, N03-02-17123,
06-02-16843, and N09-02-01136), State program 'Astronomy' (Russia),
CONACYT research grant 39560-F and 54480 (M´exico) and the Ministry
of Education and Science through the project Astrophysical
Spectroscopy of Extragalactic Objects. We thank the anonymous
referee for useful suggestions that improved the clarity of this
manuscript.}

\References

\refb Chen  K., Halpern J., 1989, ApJ, 344, 115
\refb Collier S., Crenshaw D. M., Peterson, B. M et al., 2001, ApJ, 561, 146
\refb Crenshaw D. M, Kraemer  S. B., Boggess A. et al., 1999, ApJ, 516, 750
\refb Crenshaw D. M., Kraemer, S. B., Turner T. J., Collier S., Peterson B. M., et al., 2002, 566, 187
\refb Edelson  R., Turner T. J., Pounds K. et al., 2002, ApJ, 568, 610
\refb Eracleous  M., Halpern J., 1994, ApJSS, 90, 30
\refb Halpern J. P., Eracleous M., Filippenko A. V., Chen K., 1996, ApJ, 464, 704
\refb Ho  L. C., Rudnick G., Rix H. W., et al, 2000, ApJ, 541, 120
\refb Ili\'{c} D., 2007, SerAJ, 175, 15
\refb Jovanovi\'{ c} P., Popovi\' {c} L. {\v C}. et al., 2010, ApJ, 718, 168
\refb Kollatschny  W. and Dietrich  M. 1997, A\&A, 323, 5
\refb   Kova\v {c}evi\'{c} J., Popovi\'{c} L. \v {C}, Dimitrijevi\'{c} M. S., 2010, ApJS, 189, 15
\refb Matsumoto  C., Leighly  K. M., Marshall H. L., 2001,
In X-ray Emission from Accretion onto Black Holes, Proceedings of a joint workshop held
by the Center for Astrophysics (Johns Hopkins University) and the Laboratory for High
Energy Astrophysics (NASA/ Goddard Space Flight Center) in Baltimore, MD, June 20-23, 2001, Eds.: T. Yaqoob and J. H. Krolik
570
\refb Peterson B. M., 1993, PASP, 105, 207
\refb Peterson B. M., 1999, ASPC, 175, 49
\refb Peterson B. M., Berlind P. et al., 2002,  ApJ, 581, 197
\refb Shapovalova A. I, Doroshenko V.T. et al., 2004, A\&A, 422, 925;
\refb Popovi\' {c} L. {\v C}., Shapovalova A. I., et al, 2008, PASJ, 60, 1
\refb Popovi\' {c} L. {\v C}., Shapovalova A. I., et al, 2011, A\&A, 528, 130
\refb Pounds K., Edelson, R.,  Markowitz A., Vaughan S., 2001, ApJ, 550, L15
\refb Shapovalova A. I.,Popovi\' {c} L. {\v C}., et al, 2008, A\&A, 486, 99
\refb Shapovalova A. I.,Popovi\' {c} L. {\v C}., et al, 2010a, A\&A, 509, 106
\refb Shapovalova A. I.,Popovi\' {c} L. {\v C}., et al, 2010b, A\&A, 517, 42
\refb Shemmer O., Romano  P., Bertram  R., Brinkmann  W., et al, 2001, ApJ, 561, 162
\refb Stauffer J., Schild R., Keel  W, 1983, ApJ, 270, 465
\refb Sulentic J. W. et al., 2000, ARA\&A, 38, 521
\refb Wanders I., Peterson B. M., 1996, ApJ, 466, 174

\end{document}